\documentstyle[aps,epsfig]{revtex}
\newcommand {\be}{\begin{equation}}
\newcommand {\ee}{\end{equation}}
\newcommand {\bey}{\begin{eqnarray}}
\newcommand {\eey}{\end{eqnarray}}

\begin{document}
\draft
\title{Correlation functions for a Bose-Einstein condensate in the Bogoliubov
approximation.}
\author{A.~Montina,  and E.~Arimondo}
\address{Dipartimento di Fisica and Istituto Nazionale Fisica della
Materia, Universit\`a di Pisa,
Via F. Buonarroti 2, 56127 Pisa, Italy}
\date{\today}
\maketitle
\begin{abstract}
In this article we introduce a differential equation for the first order
correlation function $G^{(1)}$ of a Bose-Einstein condensate at $T=0$.
The Bogoliubov approximation is used.
Our approach points out directly the dependence on the physical parameters.
Furthermore it suggests a numerical method to calculate $G^{(1)}$ without
solving an eigenvector problem.
 The $G^{(1)}$ equation is generalized to the case of non zero
 temperature.

\end{abstract}

\section{Introduction}
The  observations of atomic Bose-Einstein condensate (BEC) in dilute
atomic gases have triggered
a great theoretical interest for this particular state of
matter\cite{varenna}.
The Bose-Einstein condensate is a good opportunity to apply theoretically and
verify experimentally the concepts of the quantum mechanics. In fact
several
interesting theoretical features, as macroscopic quantum tunneling and
macroscopic quantum coherence, could be observed in  BEC's in the
near future.  A
condensate is characterized by a macroscopic occupation of a single
particle state and by a large spatial correlation for the atomic
spatial distribution. The long range spatial order has been studied in
a series of theoretical
papers\cite{theocohe,hoston,narasch,wong,Rohrl,castin,Glauber}. On the
experimental side interference experiments involving sodium and rubidium
condensates \cite{expecohe,hansch} have demonstrated the presence of
long-range order. 
The excellent agreement between the experimental
results and  theoretical analyses \cite{Rohrl} has confirmed the presence of
that long range order. More recent experiments have explored some features
of second order\cite{Ketterle}
and third-order\cite{Burt} atomic coherences . In ref.\cite{Ketterle} the
relationship
between the second order coherence and the interaction energy has been studied,
infering that  release energy measurements are consistent with
an unitary value  for the second order coherence of a pure condensate.
Burt et al.\cite{Burt} have measured the three-body rubidium recombination rate
of a condensate and of a cold noncondensate. They derive that the ratio of
the third order coherences in those systems is $7.4\pm2.6$, in good
agreement with the predicted value of 6.

Two standard tools to study the condensate are the
Gross-Pitaevskii equation and the Bogoliubov approximation. Because in
this approximation the hamiltonian is quadratic in the field, each property
of the system is derivable by the mean field $\psi(\vec x)$ and the first
order
correlation function $G^{(1)}(\vec x,\vec y)$, that is related to the first
order coherence properties of a condensate. The mean field is
described by the Gross-Pitaevskii equation. An extensive theoretical study
of coherence properties of BEC has been
performed by M.~Naraschewski and R.~J.~Glauber\cite{Glauber}. To calculate
the correlation functions they use the local density approximation, that is
suitable for large enough systems. Furthermore they assume that the condensate
kinetic energy is much smaller than the interaction energy. This condition is
not fullfilled in a region close to the surface of the condensate, where the
laplacian of the wave function, and therefore the kinetic energy, is not small.
A standard way to calculate the correlation functions is to solve an
eigenvector problem. For instance this method was used in\cite{Java} for
a spherically symmetric harmonic-oscillator trap to evaluate the number
of noncondensate atoms. In the anisotropic tridimensional case it is
essential to choose a suitable set of functions to reduce the dimension of
the matrix to be diagonalized. Often it is not easy to find
this set and the matrix becomes very large for the numerical
calculations, for example in the case of a
double well trap.

Purpose of the present work is to find  also for $G^{(1)}(\vec x,\vec y)$
a differential equation, similar to the Gross-Pitaevskii equation for
$\psi(\vec x)$,  in order to provide the dependence of the first
order correlation
on the physical parameters. This equation suggests an alternative method to
evaluate the correlation function.
 We introduce a differential equation for the $2\times2$ matrix
$F(\vec x,\vec y)=
\langle \vec x|F|\vec y\rangle$, with $\vec{x} $ and $\vec y$
positions in the phase space. We find that
the knowledge of $F(\vec x,\vec y)$  allows us to evaluate the correlation
functions.
The complete calculation of $G^{(1)}(\vec x,\vec y)$ is
not much more efficient than the eigenvector evaluation. However our equation
for $F(\vec x,\vec y)$  allows
to obtain easily $G^{(1)}(\vec x,\vec y)$ for a fixed $\vec y$
or integrating it with a weight function $P(\vec y)$. Our method is
very useful
if  a complete information is not required. Moreover it is suitable  to
test numerically the approximations introduced with other methods of
solutions.
At first we will consider the case of zero temperature, then we generalize
our equation to the
case of non zero temperature~\cite{nota3}.

\section{Correlation function}
In the Bogoliubov approximation\cite{bogo} the quantum boson field
$\hat\psi(\vec x)$ is written in the following way\cite{Fetter}
\begin{equation}
\label{bogo}
\hat\psi(\vec x)=\psi(\vec x)+\sum_{\lambda=1}^{\infty}[u_\lambda(\vec x)\hat
a_{\lambda}+v_\lambda^*(\vec x)\hat a_\lambda^{\dag}]
\end{equation}
where $\psi$ is determined by the time-independent Gross-Pitaevskii equation
\be
-\frac{\hbar^2}{2m}\nabla^2\psi + g|\psi|^2\psi=\mu\psi.
\ee
and $\mu$ is the chemical potential. In Eq.~(\ref{bogo}) $\hat a_\lambda$
and $\hat a_\lambda^\dagger$ are
annihilation and creation operators and $(u_\lambda,v_\lambda)$ are
the solutions of the following eigenvector
problem
\bey\label{boeq1}
{\cal L_\epsilon}u_\lambda+g\psi^2v_\lambda=E_\lambda u_\lambda  \\
\label{boeq2}
{\cal L_\epsilon}v_\lambda+g(\psi^{*})^2u_\lambda=-E_\lambda v_\lambda
\eey
with ${\cal L}_\epsilon=-\frac{\hbar^2}{2m}\nabla^2 + 2g|\psi|^2+
V-\mu+\epsilon$.
$\epsilon$ is a positive infinitesimal number that we introduce to
eliminate some divergences to be met with. To simplify the notation we
do not indicate the dependence of the eigenvectors and the eigenvalues on
$\epsilon$. The zero energy eigenvector ($\lambda=0$) is excluded in summation
of Eq.~(\ref{bogo})(as applied for instance in ref.~\cite{Java}).

$(u_\lambda,v_\lambda)$ satisfy the orthonormality and completeness relations
\be\label{orthonorma}
\int[u_\lambda(\vec x) u_{\lambda'}^*(\vec x)-
v_\lambda(\vec x) v_{\lambda'}^*(\vec x)]d^3x=\delta_{\lambda,\lambda'}
\;\;,\;\;\forall\lambda,\lambda'\geq0
\ee
\be\label{complete}
\sum_{\lambda=0}^\infty [u_\lambda(\vec x)u_\lambda^*(\vec y)-
v_\lambda(\vec x)v_\lambda^*(\vec y)]=\delta(\vec x-\vec y)
\ee

We point out that for $\epsilon=0$  $(u_0=\psi,v_0=-\psi^*)$ is the
energy eigenvector with eigenvalue $E_0=0$.
 $(u_0,v_0)$
is not normalizable, because  $\int[|u_0|^2-|v_0|^2]d^3x=
\int[|\psi|^2-|\psi|^2]d^3x=0$.

The ground state is defined by the Eqs. $\hat a_\lambda|0>=0$,
$\forall\lambda\geq1$. Explicitely using Eq.~(\ref{bogo}) we find that the
first order correlation
function for temperature $T=0$ is given by
$$
<\psi^\dag(\vec x)\psi(\vec y)>=\psi^*(\vec x)\psi(\vec y)+
\lim_{\epsilon\rightarrow0^+}\sum_{\lambda,\lambda'=1}^\infty
<0|[u_\lambda^*(\vec x)\hat a_\lambda^\dag+v_\lambda(\vec x)\hat a_\lambda]
[u_\lambda(\vec y)\hat a_\lambda+v_\lambda^*(\vec y)\hat a_\lambda^\dag]|0>
$$
\be\label{corr}
=\psi^*(\vec x)\psi(\vec y)+C(\vec x,\vec y)
\ee
With
\be\label{Co}
C(\vec x,\vec y)=\lim_{\epsilon\rightarrow0^+}
\sum_{\lambda=1}^\infty v_\lambda(\vec x)v_\lambda^*(\vec y).
\ee
Eq.~(\ref{bogo}) cannot be considered a operator identity and, to be more
rigorous,
we should have followed the Gardiner's approach~\cite{Gardi}. However the
resulting Eqs.~(\ref{corr},\ref{Co}) are not changed.
The set of Eqs.~(\ref{boeq1},\ref{boeq2},\ref{corr},\ref{Co}) defines
completely our problem.

Our first purpose is to find for $C(\vec x,\vec y)$ a compact equation,
where no eigenvector set appears and the dependence on the physical parameters
is more evident.
We introduce the annihilation operator field
\be\label{phi}
\hat\phi(\vec x)=\sum_{\lambda=0}^{\infty}\left[u_\lambda(\vec x)
\hat a_\lambda+v_\lambda^*(\vec x)\hat a_\lambda^\dag\right],
\ee
where the summation is performed over all the eigenvectors.
$\hat\phi(\vec x)$ satisfies the usual commutation relations
\be\label{commutation}
[\hat\phi(\vec x),\hat\phi^\dag(\vec y)]=\delta(\vec x-\vec y)
\ee
We then consider the state $|\tilde 0>$ defined by the equations
$a_\lambda|\tilde 0>=0$, $\forall\lambda\geq 0$.
It is evident that the Wigner function for $|\tilde 0>$ is
\be\label{wigner1}
W(\{a_\eta\},\{a_\eta^*\})\propto e^{-2\sum_{\lambda=0}^\infty a_\lambda^* a
_\lambda}
\ee
By the orthonormality relations (\ref{orthonorma}) the
Wigner function becomes
$$
W(\{a_\eta\},\{a_\eta^*\})\propto e^{-2\int d^3x\sum_{\lambda,\lambda'=0}
^{\infty}(u_{\lambda'}u_\lambda^*-v_{\lambda'}
v_\lambda ^*)a_\lambda^*a_{\lambda'}}
$$
\be\label{wigner2}
=e^{-\int d^3x\sum_{\lambda,\lambda'=0}^{\infty}\left[(u_{\lambda}^*a_\lambda
^*+v_\lambda a_\lambda)(u_{\lambda'}a_{\lambda'}-v_{\lambda'}^*
a_{\lambda'}^*)+(u_{\lambda}^*a_\lambda
^*-v_\lambda a_\lambda)(u_{\lambda'}a_{\lambda'}+v_{\lambda'}^*
a_{\lambda'}^*)\right]}
\ee
It is useful to write $a_\lambda$ as a two component vector and $u_\lambda$,
$v_\lambda$ as
$2\times2$ matrices. We will use the notations
$$
\vec a_\lambda=\left(\begin{array}{c}
Re[a_\lambda] \\
Im[a_\lambda]
\end{array}\right)
$$
$$
{\bf u}_\lambda=\left(\begin{array}{cc}
Re[u_\lambda] & -Im[u_\lambda] \\
Im[u_\lambda] & Re[u_\lambda]
\end{array}\right),\;
{\bf v}_\lambda=\left(\begin{array}{cc}
Re[v_\lambda] & -Im[v_\lambda] \\
Im[v_\lambda] & Re[v_\lambda]
\end{array}\right)
$$
\bey
\vec a_\lambda^*={\hat\sigma}_3\vec a_\lambda
\;\;,\;\;\;\;{\bf u}^*={\hat\sigma}_3{\bf u}_\lambda
\;\;,\;\;\;\;{\bf v}_\lambda^*={\hat\sigma}_3{\bf v}_\lambda
\eey
where ${\hat\sigma}_3$ is the Pauli matrix with the diagonal elements $1$ and
 $-1$. With this vector and matrix notations Eq.~(\ref{wigner2}) becomes
\be
W(\{\vec a_\eta\})\propto e^{-2\int d^3x\sum_{\lambda,\lambda'=0}^{\infty}
(\vec a_\lambda^\dag {\bf u}_{\lambda}^\dag
+\vec a_\lambda^{*\dag} {\bf v}_\lambda^{*\dag})({\bf u}_{\lambda'}
\vec a_{\lambda'}-{\bf v}_{\lambda'}^*
\vec a_{\lambda'}^*)}
\ee
 Eqs.~(\ref{boeq1},\ref{boeq2}) allow us to find that
\bey\label{scambio1}
H_1({\bf u} \vec a+{\bf v}^*\vec a^*)=E_\lambda({\bf u} \vec a-{\bf v}^*
\vec a^*) \\
\label{scambio2}
H_2({\bf u} \vec a-{\bf v}^*\vec a^*)=E_\lambda({\bf u} \vec a+{\bf v}^*
\vec a^*)
\eey
where $H_1={\cal L}_\epsilon+g\Psi^2\hat\sigma_3$,
$H_2={\cal L}_\epsilon-g\Psi^2\hat\sigma_3$ and
$\Psi$ is a $2\times2$ matrix constructed by $\psi$ as the $\bf u$ and $\bf v$
matrices.
>From these equations we deduce
\be\label{dia}
(H_1\cdot H_2)^{1/2}({\bf u}\vec a-{\bf v}^*\vec a^*)=E_\lambda({\bf u}\vec a-
{\bf v}^*\vec a^*)
\ee
and from Eqs.~(\ref{scambio1},\ref{dia})
\be
W(\{\vec a_\eta\})\propto  e^{-2\int d^3x\sum_{\lambda,\lambda'=0}^{\infty}
(\vec a_\lambda^\dag {\bf u}_{\lambda}^\dag
+\vec a_\lambda^{*\dag} {\bf v}_\lambda^{*\dag} )(H_1\cdot H_2)^{-1/2}H_1(
{\bf u}_{\lambda'}\vec a_{\lambda'}+{\bf v}_{\lambda'}^*\vec a_{\lambda'}^*)}.
\ee
Note that if we did not use our real notation we had to introduce antilinear
operators.

We now perform the transformation
\be
\vec\phi(\vec x)=\sum_{\lambda=0}^{\infty}\left[{\bf u}_{\lambda}(\vec x)
\vec a_{\lambda}+{\bf v}_{\lambda}^*(\vec x)\vec a_{\lambda}^*\right]
\ee
to obtain the $W$ as a function of the field $\vec\phi(\vec x)$ that
corresponds to the quantum field $\hat\phi(\vec x)$ of Eq.~(\ref{phi})
\be
W(\{\vec\phi\})\propto  e^{-2\int d^3x
\vec\phi^\dag(\vec x)(H_1\cdot H_2)^{-1/2}H_1\vec\phi(\vec x)}
\ee
It is evident that $(H_1\cdot H_2)H_1=
H_1(H_2\cdot H_1)$. Therefore $M=(H_1\cdot H_2)^{-1/2}H_1=
H_1(H_2\cdot H_1)^{-1/2}=M^\dag$ , {\it i.e.}
$M$ is a symmetric operator.
More in general
\be\label{scambio}
f(H_1\cdot H_2)\cdot H_1=H_1\cdot f(H_2\cdot H_1)
\ee
It is then easy to demonstrate that the mean weighted with the Wigner
function is given by
\be\label{Fcorr}
<\phi_i(\vec x)\phi_j(\vec y)>_W=\frac{1}{4}M^{-1}_{(\vec x,i),(\vec y,j)}
\equiv\frac{1}{4}\left[H_1^{-1}(H_1\cdot H_2)^{1/2}
\right]_{(\vec x,i),(\vec y,j)}
=\frac{1}{4}\langle\vec x,i|H_1^{-1}(H_1\cdot H_2)^{1/2}|\vec y,j\rangle.
\ee
In fact, if $\hat T_{k,l}$ is a symmetric matrix then exists a orthogonal
transformation
$z_k=\sum_l\hat O_{k,l}Z_l$ that diagonalizes $\hat T$. Therefore
$$
\int z_i z_j e^{-2\sum_{k,l}\hat T_{k,l}z_k z_l}d\vec z=
\sum_{i',j'}\hat O_{i,i'}\hat O_{j,j'}\int Z_{i'}Z_{j'}
e^{-2\sum_k\hat T'_{k,k}Z_kZ_k}d\vec Z
$$
\be
=1/4\sum_{i',j'}O_{i,i'}O_{j,j'}(\hat T'^{-1})_{i',j'}=1/4(\hat T^{-1})_{i,j}
\ee
The expectation value of an operator $F(\hat\phi,\hat\phi^\dag)$, symmetrically
ordered, is given by the mean of the classical function $F(\phi,\phi^*)$
weighed with the Wigner function, therefore
\be\label{simm}
1/2<0|\hat\phi^\dag(\vec x)\hat\phi(\vec y)+h.c|0>=
<(\phi_1(\vec x)-i\phi_2(\vec x))(\phi_1(\vec y)+i\phi_2(\vec y))>_W
\ee
Combining Eq.~(\ref{commutation},\ref{Fcorr},\ref{simm}) we
find that
\be\label{cor2}
\tilde C(\vec x,\vec y)\equiv <0|\hat\phi^\dag(\vec x)\hat\phi(\vec y)|0>=
F_{(\vec x,1),(\vec y,1)}+F_{(\vec x,2),(\vec y,2)}
+i F_{(\vec x,1),(\vec y,2)}-i F_{(\vec x,2),(\vec y,1)}
\ee
where the operator $F$ is defined by
$$
F\equiv \frac{1}{4}H_1^{-1}\left[(H_1\cdot H_2)^{1/2}-H_1\right]
$$
The $\lambda=0$ term should be subtracted from
 $\tilde C(\vec x,\vec y)$ in order  to obtain the $C(\vec x,\vec y)$
 quantity defined in
Eq.~(\ref{Co})
\be\label{sub}
C(\vec x,\vec y)=\lim_{\epsilon\rightarrow0^+}
\left[\tilde C(\vec x,\vec y)-v_0(\vec x)v_0^*(\vec y)\right]
\ee
where $\tilde C(\vec x,\vec y)$ and $v_0(\vec x)$ depend implicitly by
the parameter $\epsilon$. $v_0(\vec x)$ can be calculated solving the
dynamical equations obtained replacing $E_\lambda$ with $i\hbar
\frac{\partial}{\partial t}$ in Eqs.~(\ref{boeq1},\ref{boeq2}). In fact, if
$u(\vec x,t)$ and $v(\vec x,t)$ are the solution of these equations then
$$
v_0(\vec x)\propto \int_{-\infty}^{\infty}dt\int_{0}^{\eta}
dEe^{i E t}v(\vec x,t)
=\int_{-\infty}^{\infty}dt\frac{1}{i t}\left(e^{i\eta t}-1\right)v(\vec x,t)
$$
\be
u_0(\vec x)\propto \int_{-\infty}^{\infty}dt\int_{0}^{\eta}dE
e^{i E t}u(\vec x,t)
=\int_{-\infty}^{\infty}dt\frac{1}{i t}\left(e^{i\eta t}-1\right)u(\vec x,t)
\ee
where $\eta$ is a number such that $E_0<\eta<E_1$. It is
convenient to get $u(\vec x,0)=v(\vec y,0)=\psi$ as initial state and
to introduce a temporal gaussian window to lower the convergence time.

We have reduced our problem to the evaluation of the operator
$F$.  $F$ satisfies the following equation
\be\label{Equa}
H_1\cdot F=\frac{1}{4}\left[(H_1\cdot H_2)^{1/2}-H_1\right]\equiv S
\ee
that is
\be\label{Yukawa}
-\nabla^2F(\vec x,\vec y)+M^2F(\vec x,\vec y)=\frac{2m}{\hbar^2}
S(\vec x,\vec y)
\ee
where
\be\label{mass}
M^2=\frac{2m}{\hbar^2}\left[V+2g\Psi^*\Psi-g\Psi^2-\mu+\epsilon\right]
\ee
Eq.~(\ref{Yukawa}) is a Yukawa-like equation with a coordinate dependent
mass
and a charge distribution $S(\vec x,\vec y)$ in $\vec x$. Both S and M are
$2\times2$ matrices.

If $S$ is known, $F(\vec x,\vec y)$ can be evaluated finding for all the
$\vec y$ positions the
stationary state of the following differential equation:
\be\label{coll}
\frac{\partial}{\partial t}F(\vec x,\vec y)+H_1F(\vec x,\vec
y)=S(\vec x,\vec y).
\ee
Every function whose evolution is determined by Eq ~(\ref{coll}) collapses
in this
stationary state
because $H_1$ is a positive eigenvalue operator.

The standard method to calculate the correlation function is to solve the
eigenvector problem of Eqs.~(\ref{boeq1},\ref{boeq2}) using
Eqs.~(\ref{corr},\ref{Co}). However in some cases the matrix  to be
diagonalized becomes too large to be handled.
Eq.~(\ref{Equa}) can be useful to extract informations about the
correlation function without the resolution of an eigenvector problem.

If we want to calculate $C(\vec x,\vec y)$ with a fixed $\vec y$ or integrating
$\vec y$ with a weight function $P(\vec y)$
\be\label{ridu}
C(\vec x)=\int C(\vec x,\vec y)P(\vec y)d^3y
\ee
we have to solve only the differential equation~(\ref{coll}) with $\vec y$
fixed or with the source term
\be\label{source}
(\vec S_0)_i(\vec x)=\sum_{j=1}^{2}\int S_{i,j}(\vec x,\vec y)\vec P_j(\vec y)
d^3y.
\ee
where $\vec P(\vec y)=\left(\begin{array}{c}P(\vec y)\\P(\vec y)\end
{array}\right)$. This approach allows to
decrease considerably the computation time.

The  question  to be solved is the evaluation of $\vec S_0$. To
calculate the source term the square root in
the second term of Eq.~(\ref{Equa}) should be known, but that requires to
solve an eigenvector
problem that could be avoided through an alternative method.
It is well known from the  Dirac  theory that a square root of the operator
$-\nabla^2+m^2$ is a
local operator with first order differential derivatives that multiply
anticommuting matrices. The only difference between that square root and the
nonlocal operator $\sqrt{-\nabla^2+m^2}$ is the sign of the eigenvalues,
that in the last case are all positive.

A local non-positive square root exists also for $R=H_1\cdot H_2$.
$R$ has the form, apart from a constant factor,
\be
R=\left[-\nabla^2+Q_1+Q_2\hat\sigma_1+Q_3\hat\sigma_3\right]
\left[-\nabla^2+Q_1-Q_2\hat\sigma_1-Q_3\hat\sigma_3\right]
\ee
where $Q_1$, $Q_2$ and $Q_3$ are three real functions.
It is easy to demonstrate that the operator
\be\label{diraclike}
H_r=\hat\sigma_2(-\nabla^2+Q_1)-i\hat\sigma_3Q_2+i\hat\sigma_1Q_3
\ee
is square root of R, that is $H_r^2=R$.

If $\vec P_+$ and $\vec P_-$ are the projections of $\vec P$
respectively over the positive and negative eigenvalue subspaces of $H_r$
then
\be
\vec S_0(\vec x)=H_r(\vec P_+-\vec P_-)
\ee
We now describe how to handle $\vec P_+-\vec P_-$.
If $\vec P(\vec x,\tau)$ is solution of the equation
\be\label{evoe}
i\frac{\partial}{\partial\tau}\vec P=H_r \vec P
\ee
it is evident that
$$
\int_0^\infty dE\int_{-\infty}^\infty \vec P(\vec x,\tau)e^{iE\tau}d\tau=
\vec P_+(\vec x)
$$
\be\label{intwin}
\int_{-\infty}^0 dE\int_{-\infty}^\infty \vec P(\vec x,\tau)e^{iE\tau}d\tau=
\vec P_-(\vec x)
\ee
Performing the integration in E, we obtain
\be\label{sorg}
\vec P_+(\vec x)-\vec P_-(\vec x)=\frac{1}{\pi}\lim_{\eta\rightarrow0^+}
\int_\eta^\infty\frac{1}{\tau}[\vec P(\vec x,\tau)-\vec P(\vec x,-\tau)]d\tau
\ee
Therefore
\be\label{sorg2}
\vec S_0(\vec x)=\frac{1}{\pi}H_r \lim_{\eta\rightarrow0^+}
\int_\eta^\infty\frac{1}{\tau}[\vec P(\vec x,\tau)-\vec P(\vec x,-\tau)]d\tau
\ee
The solution of Eq.~(\ref{evoe}) allows to evaluate the source term of
Eq.~(\ref{source}). The direct calculation of
Eq.~(\ref{sorg}) is probably not
the best choice. In fact, if the ratio between the greatest and lowest
frequencies is too large then the integration step has to be too small
with respect to the integration time. In this case it is convenient to perform
the energy integration of Eqs.~(\ref{intwin}) over the windows
$(E_1,E_2)$, $(E_2,E_3)$, $(E_3,E_4)$, $...$, with $E_1>E_2>E_3...$ and
to choose for each window a suitable integration step. It is also
convenient to use a temporal gaussian window to reduce the calculation time.
In this article we do not discuss these numerical questions into details.

\section{Numerical tests}
At this stage we have all the tools to evaluate the
correlation function of Eq.~(\ref{ridu}).
We have checked numerically in the one-dimensional case that the same
source terms
are obtained from
Eq.~(\ref{sorg2}) and by the diagonalization. Also the validity
of Eq.~(\ref{Equa}) has been verified numerically.

\begin{figure}
\epsfig{figure=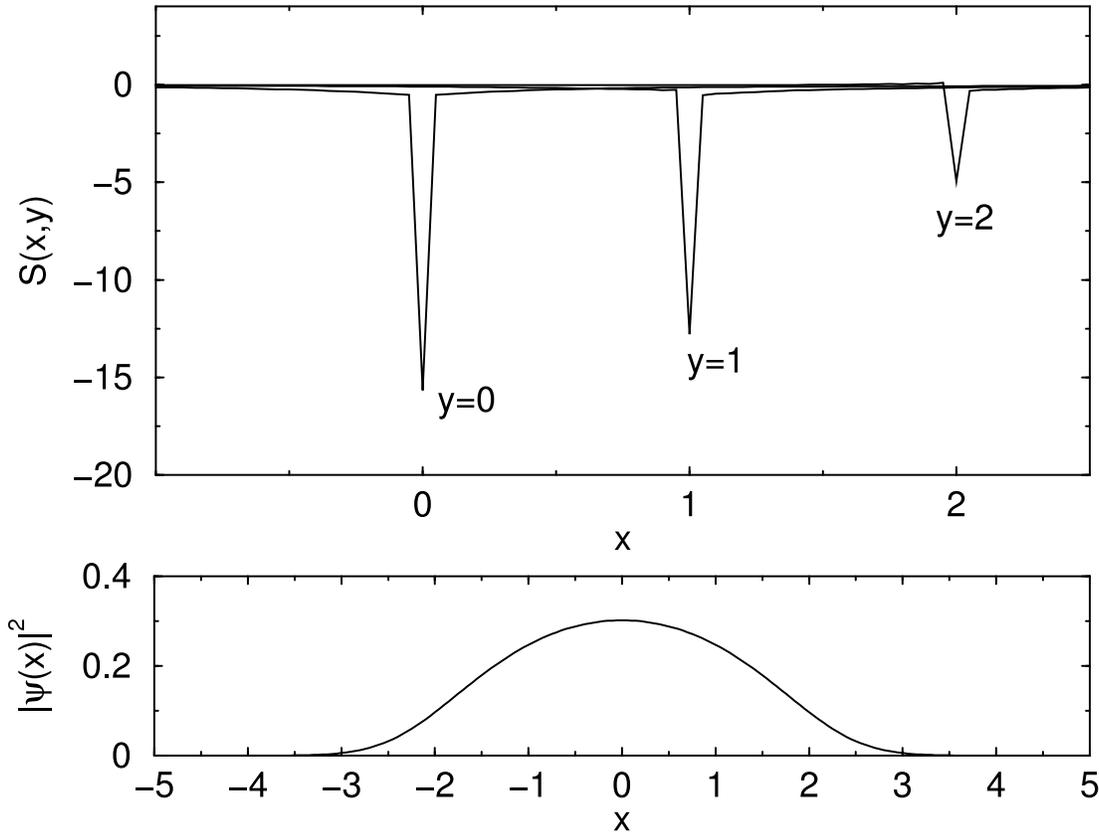,width=11cm,angle=-90} 
\caption[]{In (a) source term $S_{1,1}(x,y)$ in the one-dimensional case as a
function of $x$, calculated for three
values of $y$ and a coupling constant $g=10$. In (b) density
function $|\psi(x)|^2$ from the Gross-Pitaevskii equation for the same
parameters.
Adimensional unities are used.}
\label{fig1}
\end{figure}

In order to test the technique we have considered the case of a
one-dimensional harmonic
trap with $V(x)=1/2x^2$ and a coupling constant $g=10$ ($\hbar=m=1$).
The $\psi(x)$ solution of the Gross-Pitaevskii equation is reported in
the lower part of Fig.~\ref{fig1}.  Instead sections of $S_{1,1}( x, y)$
 are reported in the upper part of the
figure for different
values of $y$. The results of the figure point out  that the source term
is a near diagonal operator.
In fact $S$ is a sum of
a diagonal matrix and a smooth function $f(x,y)$.
In other terms, for every $y$, the source has a point-like charge
with a cloud around.

We have considered also a tridimensional case.
The studied system is constituted by $^{87}Rb$ atoms confined in a spherical
harmonic trap in the $|F=1,m_f=-1\rangle$ hyperfine sublevels. For the
scattering length we have used $a=109.1$ a.u.. The trap frequency is
supposed
$\omega=2\pi\cdot300$ s$^{-1}$.
We have imposed $\vec y=0$ in Eq.(32) in order to exploit the trap symmetry
and therefore to
simplify the calculation. The application into the case of asymmetric
trap require only some additional algebra\cite{nota2}.  In Fig.~2 we plot 
$C(r)=C(\vec
x,\vec y=0)$ as a
function of $r=|\vec x|$ for different values of the boson number $N$. The
functions
obtained by diagonalization and solving our differential equation overlap and
therefore are indistinguishable in the plot.
\begin{figure}
\epsfig{figure=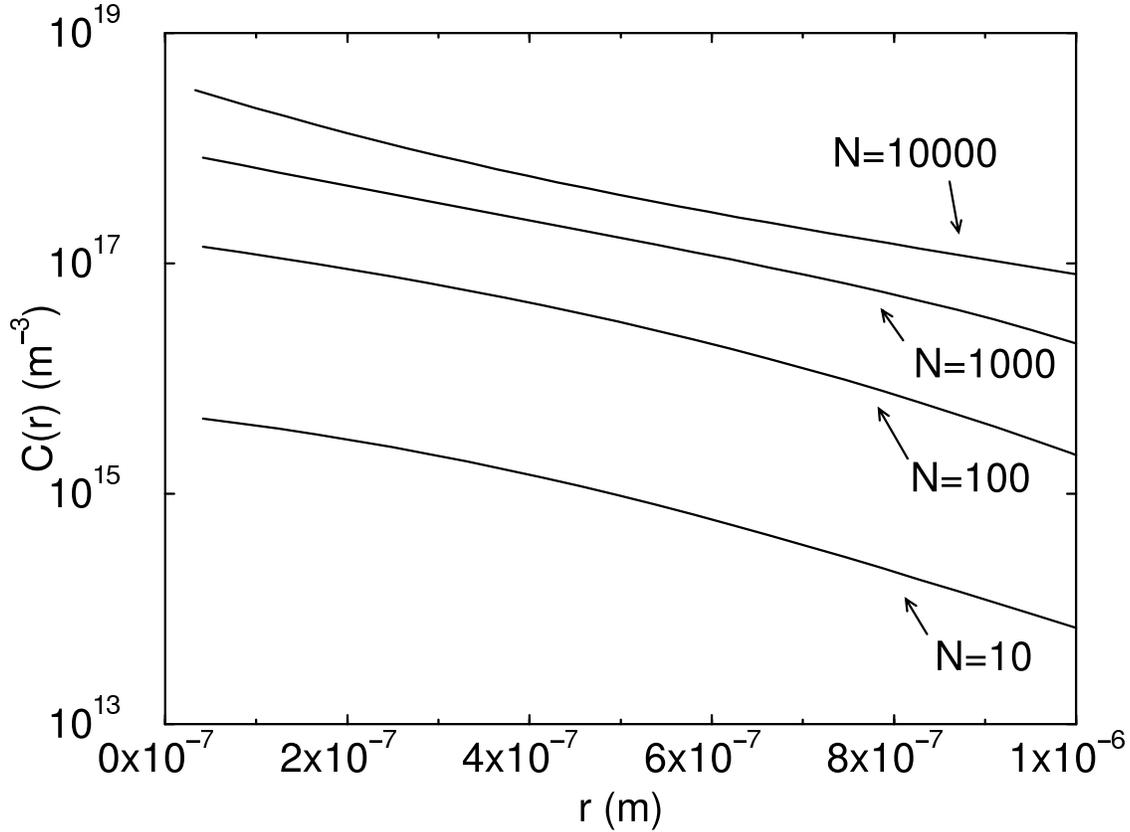,width=11cm,angle=-90} 
\caption[]{Plot of $C(r)=C(\vec x,\vec y=0)$ as a function of $r=|\vec x|$
for some values of the boson number $N$. We have considered
$^{87}Rb$ atoms trapped in the $|F=1,M_m=-1\rangle$ hyperfine sublevel.}
\label{fig2}
\end{figure}

We note that $C(r=0)$ increases with $N$. This is obvious because the correlations
of the field fluttuations are a consequence of the Gross-Pitaevskii non-linear term.
The variation scale of $C(r)$ is given by $1/M(r)$ and for $N=0$ its magnitude is of
the order of $sqrt{\hbar/m\omega}=6.2\cdot10^{-7}m$.

\section{Finite temperature}
Eq.~(\ref{Equa}) can be generalized to include the finite temperature
fluctuations. Eq.~(\ref{wigner1}) is replaced by
\be
W(\{a_\eta\},\{a_\eta^*\})\propto e^{-\sum_{\lambda=0}^\infty
\frac{a_\lambda^* a_\lambda}{1/2+\left(e^{\beta_\lambda}-1\right)^{-1}}}
\ee
where $\beta_\lambda=E_\lambda/kT$. For $T\gg E_\lambda$ the correct classical
distribution is obtained.

Using Eq.~(\ref{dia}) we find $W$ as a function of $\vec\phi$
\be
W(\{\vec\phi\})\propto  e^{-2\int d^3x
\vec\phi^\dag(\vec x)(H_1\cdot H_2)^{-1/2}A_T^{-1}H_1\vec\phi(\vec x)}
\ee
where $A_T$ is the operator
\be
A_T=1+2\left(e^{\frac{(H_1\cdot H_2)^{1/2}}{kT}}-1\right)^{-1}
\ee
The $F$ operator of Eq.~(\ref{cor2}) is replaced by
\be\label{Tfin}
F_T\equiv \frac{1}{4}H_1^{-1}\left[A_T\cdot(H_1\cdot H_2)^{1/2}-H_1\right]
\ee
It is possible to find a relation between $F_T$ and $F=F_0$. We subtract
$1/4$ from the two terms of Eq.~(\ref{Tfin}) and
multiply them by $A_T^{-1}$.
We obtain using Eq.~(\ref{scambio})
\be
H_1\cdot B_T^{-1}(F_T-1)=\frac{1}{4}\cdot(H_1\cdot H_2)^{1/2}.
\ee
where
\be\label{EquaT}
B_T=1+2\left(e^{\frac{(H_2\cdot H_1)^{1/2}}{kT}}-1\right)^{-1}.
\ee
Therefore
\be
B_T^{-1}(F_T-1)=B_{T'}^{-1}(F_{T'}-1)
\ee
Setting $T'=0$ we finally find
\be\label{rela}
F_T=1+B_T\cdot(F-1).
\ee
We can perform the following expansion
$$
B_T=1+2e^{-\frac{(H_2\cdot H_1)^{1/2}}{kT}}\left(1-
e^{-\frac{(H_2\cdot H_1)^{1/2}}{kT}}\right)^{-1}
$$
\be
=1+2\sum_{n=1}^{\infty}e^{-\frac{n(H_2\cdot H_1)^{1/2}}{kT}}.
\ee
If $\vec F_i(\vec x)=\sum_j\int F_{i,j}(\vec x,\vec y)\vec P_j(\vec y)d^3y$ and
$\vec F_T(\vec x)=\sum_j\int (F_T)_{i,j}(\vec x,\vec y)\vec P_j(\vec y)d^3y$
then
$$
\vec F_T=\vec F+2\sum_{n=1}^{\infty}e^{-\frac{n(H_2\cdot H_1)^{1/2}}{kT}}
(\vec F+\vec P)
$$
\be
=\vec F+2\sum_{n=1}^{\infty}e^{-\frac{n\tilde H_r}{kT}}
(\vec F_++\vec P_+)+2\sum_{n=1}^{\infty}e^{\frac{n\tilde H_r}{kT}}
(\vec F_-+\vec P_-)
\ee
where $\vec F_+$, $\vec P_+$ and $\vec F_-$, $\vec P_-$ are the projections
over the positive and negative eigenvalues subspaces of $\tilde H_r=\hat
\sigma_3 H_r\hat\sigma_3$.
If $\vec F_\pm(\vec x,\tau)$ and $\vec P_\pm(\vec x,\tau)$ are the solution of
the differential equation~\cite{nota}
\be\label{diffeq2}
\frac{\partial}{\partial\tau}(\cdot)=\tilde H_r (\cdot)
\ee
with $\vec F_\pm(\vec x,0)=\vec F_\pm$ and
$\vec P_\pm(\vec x,0)=\vec P_\pm$, then
\be
\vec F_T(\vec x)=\vec F(\vec x)+2\sum_{n=1}^{\infty}\left[(\vec F_++\vec P_+)
(\vec x,-\frac{n}{kT})+(\vec F_-+\vec P_-)(\vec x,\frac{n}{kT})\right]
\ee
Equations similar to the ~(\ref{cor2},\ref{sub}) ones can be defined
for temperatures $T$ different from zero.
Therefore we have shown that it is possible derive the correlation function
for $T\ne0$ by knowing it for $T=0$.

\section{conclusion}
In conclusion we have introduced a differential equation that is useful to
evaluate
numerically the first order correlation function $G^{(1)}(\vec x,\vec y)$
for some fixed $\vec y$ or its integration over $\vec y$ with a weight
function
$P(\vec y)$.
In the Bogoliubov approximation the hamiltonian is quadratic in the field,
therefore the ground state is a squeezed state, that is, the Wigner function
is a  gaussian one for all its infinite modes. The gaussian parameters are
$\psi(\vec x)$, that defines its position in phase space, and the function $F
(\vec x,\vec y)$ that we have introduced.
There are not other free parameters. Therefore each property of the BEC is
derivable by solving the Gross-Pitaevskii equation and Eq.~(\ref{Equa}).
In particular the resolution of these equations allows to evaluate the higher
order correlation functions.

Our method can be applied to study the quantum fluttuations of the condensate
in double-well traps, improving the two mode model, that is the standard
approach to deal with these problems \cite{Jack}.


\begin{references}
\bibitem{nota}  To obtain $\vec F_\pm$ and $\vec P_\pm$ we can use equations
(\ref{intwin}). However there could be a problem in solving numerically
Eq.~(\ref{diffeq2}). 
The inevitable residual components $\vec F-\vec F_\pm$ and $\vec P-\vec P_\pm$
grow in the course of the time integration. It is then necessary to clean up
these components at suitable integration times $\Delta t$, $2\Delta t$, ... .
If the evaluation time becomes too large, an alternative numerical approach
should be used.
\end{references}
\end{document}